\documentstyle[12pt]{article}
\newcommand{\beq}{\begin{equation}}
\newcommand{\bea}{\begin{eqnarray}}
\newcommand{\eeq}{\end{equation}}
\newcommand{\eea}{ \end{eqnarray}}
\newcommand{\nn}{\nonumber}
\newcommand{\el}{&&\nonumber\\}

\newcommand{\psm}{pseudo-SU(3) model}
\newcommand{\psma}{pseudo-SU(3) model }

\newcommand{\al}{\alpha}

\newcommand{\la}{\lambda}

\newcommand{\k}{\kappa}
\newcommand{\lao}{\lambda_1}
\newcommand{\muo}{\mu_1}
\newcommand{\ko}{\kappa_1}
\newcommand{\lat}{\lambda_2}
\newcommand{\mut}{\mu_2}
\newcommand{\kt}{\kappa_2}

\newcommand{\sutrqn}[4]{(#1,#2) \,#3 \,#4}

\newcommand{\sutrcc}[3]{\langle \, #1 ;\, #2\, ||\, #3
\,\rangle}

\newcommand{\lmd}{$(\lambda, \, \mu)$}
\newcommand{\lmda}{$(\lambda, \, \mu)$ }
\newcommand{\lm}{(\lambda, \, \mu)}

\renewcommand{\thefootnote}{\fnsymbol{footnote}}
\begin{document}
\newfont{\dtt}{cmr12}
\begin{center}
{\Large \bf
Generalized Pseudo-SU(3) Model and
Pairing\footnote{Work
supported in part by a grant from the U.S. National
Science Foundation.}
}
\\[1 cm]
D. Troltenier and C. Bahri
\\[0.3 cm]
\,{\sl Department of Physics and Astronomy,
Louisiana State University}\\
  {\sl Baton Rouge, LA 70803--4001, U.S.A.}
\\[0.8 cm]
J. P. Draayer\footnote{Supported by the Deutsche
Forschungsgemeinschaft (DFG).
On Leave from the Department of Physics and
Astronomy, Louisiana State University,
Baton Rouge, LA 70803-4001, U.S.A.}
\\[0.3 cm]
\,{\sl Institut f\"{u}r Theoretische Physik,
Universit\"{a}t T\"{u}bingen}\\
  {\sl Auf der Morgenstelle 14, D-72076
T\"{u}bingen, Germany}
\\[2 cm]
\end{center}

\begin{abstract}
The \psma is extended to explicitly include the spin and
proton-neutron
degrees of freedom.
A general formalism for evaluating matrix elements of
one-body and two-body
tensor operators within this framework is presented.
The pairing interaction, which couples different
irreducible representations
of SU(3), is expressed in terms of pseudo-space
tensors and a general result
is given for calculating its matrix elements.
The importance of pairing correlations in \psma
calculations is demonstrated
by examining the dependence of wavefunctions, low-
energy collective excitation
spectra, and moments of inertia on the strength of the
pairing interaction.
\end{abstract}

\renewcommand{\thefootnote}{\arabic{footnote}}
\setcounter{footnote}{0}

\section{Introduction}

Over the past few years there have been a number of
discoveries in
nuclear structure physics (superdeformed shapes,
identical bands,
scissors mode, etc.) which have raised a number of
interesting
questions, many of which remain unanswered.
In addition, one can anticipate the identification of
additional new
phenomena in the near future due to ongoing
improvements and innovations
in experimental techniques (4$\pi$ detectors,
radioactive beams, etc.).
These experimental developments seem not to be
matched by a comparable
record of theoretical achievements focused on the
questions that are being
raised.

However, it is commonly accepted that the nuclear
shell model should
be able to address these issues and provide answers to
these questions.
The problem is that most shell-model theories are
limited by the large
dimensionalities of the required model spaces.
For example, existing algorithms and available
computing equipment still
only allow for reasonably complete standard shell-
model calculations
in the mass range $A \le 28$.
And even the best of these calculations do not include a
sufficient number
of multi-shell configurations to adequately address the
issues raised by
experiments that probe details of the momentum and
current distributions
in nuclei.
Furthermore, most of the new discoveries are for much
heavier nuclei,
typically in the mass range $A\ge 100$.

When faced with a situation like the one just described,
it is essential
to take full advantage of symmetries, those which are
only approximately
fulfilled as well as those that are exact.
The selection rules that are associated with such
symmetries generate,
respectively, weakly coupled and disconnected
subspaces of the full
space and this in turn allows for a significant (orders of
magnitude)
reduction in the dimensionality of the model space.
Since matrix elements of the system's Hamiltonian
between these subspaces
only vanish if the symmetry is exact, the question that
must be addressed
is whether or not using the identified approximate
symmetry is physically
justified.

Pseudo-spin \cite{AH69,HA69} is such a symmetry
and it serves as the
conceptual foundation for the subject of this
contribution, namely,
the \psma \cite{RR73}, and its extension, the pseudo-
symplectic model
\cite{TD94}.
The success of the pseudo-spin concept relies on the
fact that the
harmonic oscillator structure of the usual shell-model
Hamiltonian,
which is destroyed by the spin-orbit interaction, is
restored under
a transformation to the pseudo-space/spin picture.
The idea is to express all normal-space shell-model
quantities in terms
of their pseudo-space equivalents by applying a normal
$\leftrightarrow$
pseudo transformation to both operators and
wavefunctions \cite{CM92}.
Since for  practical purposes the harmonic oscillator
structure of the
Hamiltonian is restored in the pseudo-space scheme,
this transformation
makes it possible to invoke powerful group theoretical
truncation methods
so shell-model calculations can be carried out even for
heavy nuclei with
$A \ge 100$.

There are still questions being raised about the physical
implications of
the approximation one is making by assuming that the
pseudo-spin symmetry
is exact.
However, there have been a number of publications
\cite{RR73,BH82,BD92,TN94}
which demonstrate in both single-particle and many-
particle systems that
even if it is not an exact symmetry, for practical
applications pseudo-spin
is a sufficiently good symmetry to be of major physical
importance.
The reader is referred to these publications for
assurance regarding the
validity of the pseudo-spin concept.
Here we focus only on an algebraic many-particle
shell-model theory that
takes full advantage of pseudo-spin symmetry,
namely, the \psm.
The usefulness and predictive power of this model
have been demonstrated
in many application ranging from the calculation of
collective excitation
spectra \cite{DW84} and a scissor's mode study
\cite{CD87} to fundamental
problems likes the half-lives of $\beta\beta$-decay
modes \cite{CH94}.

In this contribution we propose extensions of the
\psma in order to widen
its range of applicability and in so doing increase the
capabilities of
the model for describing and predicting new and
interesting experimental
phenomena:
\begin{itemize}
\item The ansatz of reference \cite{CD87} is
generalized by explicitly
including spin degrees of freedom in a full proton-
neutron formulation
of the theory.
This generalization makes it possible to describe
properties of even-odd,
odd-even, and odd-odd nuclei in addition to those of
even-even ones.
Within this extended approach, tensor operators are
defined and
expressions for calculating their matrix elements are
given.
\item In spite of many successful applications of the
\psm, the
Hamiltonian that has been used in most applications is
highly schematic
in nature and needs to be extended. The usual structure
taken for the
\psma Hamiltonian is
\beq
H = H_0 - \frac{\chi}{2} Q^a\cdot Q^a +
R_{\mbox{coll}}
\label{eq:fh}
\eeq
where
$H_0$ is the harmonic oscillator Hamiltonian, $Q^a$
denotes the algebraic
quad\-ru\-pole operator (see ref. \cite{TD94}), and
$R_{\mbox{coll}}$
stands for a residual interaction that is introduced to
describe properties
of low-energy collective excitations.
A comparison with other shell-model theories indicates
the importance of
introducing a short range interaction, and therefore it
seems reasonable
to add a pairing interaction to terms already included in
Eq.(\ref{eq:fh})
\cite{RS80,BS69}.
\end{itemize}

The reason the pairing interaction has not been
introduced previously is a
technical matter:
Basis states of the \psma are organized according to
irreducible
representations (irreps) of SU(3) which are labeled by
\lmd.
In the past it was not possible to calculate matrix
elements between
different SU(3) irreps (except for a few relatively
simple cases), but
recently a general code was released that removes this
limitation
\cite{BD94}.
The point is that the terms that are part of the traditional
Hamiltonian
of Eq.(\ref{eq:fh}) do not couple different SU(3)
irreps while the pairing
interaction has non-vanishing matrix elements between
states of different
\lmd. It follows from this that the calculation is far
more interesting
and challenging when the pairing is present
\cite{CB94}.

The paper is organized as follows:
In the next section the structure of basis functions and
operators of
the generalized \psma are defined and general
expressions for evaluating
one-body and two-body matrix elements are
determined.
In Section 3, the pairing interaction is expanded in
terms of SU(3)
tensor operators and expressions for its matrix
elements are given.
The importance of pairing for the \psma is illustrated in
Section 4 by
investigating its influence on wavefunctions and
excitation spectra.
A summary is presented in Section 5.

\section{Wavefunctions, operators, and matrix
elements}
\subsection{Wavefunctions}
Wavefunctions of the \psma which take the spin and
proton-neutron degrees
of freedom explicitly into account are labeled by the
eigenvalues of the
Casimir invariants of the group chains depicted in Fig.
\ref{fig:wf}.
\\[1 cm]
{\bf Figure \ref{fig:wf}}
\\[1 cm]
Since the \psma is a many-particle theory, it is to be
understood that its
wavefunctions are subject to the Pauli Exclusion
Principle, that is, they
must be anti-symmetric under the interchange of
identical particles.
Because a  particle-permutation $\leftrightarrow$
unitary-symmetry
complementarity applies, this requirement can be met
by restricting the
basis states to the antisymmetric irreps of $U(2
\Omega_\sigma)$, where
$\sigma=\pi$ for protons and $\sigma=\nu$ for
neutrons \cite{CL93} and
where $\Omega_\sigma$ =
$(\eta_\sigma+1)(\eta_\sigma+2)/2$ is the number
of normal parity levels in the $\eta_\sigma$ shell.

These anti-symmetric irreps of $ U(2 \Omega_\sigma)$
are characterized
by $\left[ 1^{N_\sigma} \right]$ where $N_\sigma$ is
the number of normal
parity particles.
The $N_\sigma$ in this result can be determined by a
simple procedure with
the help of a Nilsson diagram (see, for example,
\cite{LS78}):
At the experimentally determined $\beta_2$ value, the
proton (neutron)
Nilsson levels (unique parity as well as normal parity)
are filled pairwise
(to accomodate the spin degree of freedom) from the
bottom of the well up
until the total number of protons (neutrons) equals the
number in the nucleus
under consideration.
If shifts in $\beta_2$ by $10$ to $20$ percent of the
original value do not
influence the distribution of particles in the unique and
normal parity
orbitals, one can assume that the $N_\sigma$
determined in this way is
more or less unique for that particular nucleus.
(A Nilsson-like single-particle code which improves
upon and automates this
procedure is currently being developed. Specifically,
for any given nucleus
it calculates  $N_\sigma$ as a function of the
deformation $\beta_2$ together
with statistical measures that characterize the sensitivity
of the $N_\sigma$
to changes in $\beta_2$.)

The irreps of $U(\Omega_\sigma)$ and
$U(2)^{(\sigma)}$ are denoted by
$[f_\sigma]$ and $[\bar{f}_\sigma ]$, respectively.
The overall antisymmetry in $U(2 \Omega_\sigma)$
and the particle-permutation
$\leftrightarrow$ unitary-symmetry complementarity
means that $[f_\sigma]$ is
related to $[\bar{f}_\sigma ]$ through row
$\leftrightarrow$ column exchange
\cite{CL93}.
Since the overall attractive nature of the nucleon-
nucleon interaction insures
that spatially symmetric irreps of the
$U(\Omega_\sigma)$ lie energetically
lowest, in this work only the most symmetric irreps of
$U(\Omega_\pi)$ and
$U(\Omega_\nu)$ are taken into account.
The physical consequence of this restriction is that no
couplings to spin-flip
excitations are considered.
This approximation can be justified by noting that spin-
flip modes normally lie
significantly higher in energy than the low-energy
collective excitations which
are of primary interest here.

The reduction $U(\Omega_\sigma) \supset SU(3) $
(see Fig. \ref{fig:wf})
determines the ($\lambda_\sigma, \mu_\sigma$) irreps
that are contained
in a given $[f_\sigma]$.
Multiple occurences of the same ($\lambda_\sigma,
\mu_\sigma$) in a fixed
$[f_\sigma]$ are distinguished by a running integer
index $\alpha_\sigma$.
The SU(3) labels of the total wavefunction are
calculated by taking all
possible products
\{($\lambda_\pi, \mu_\pi$)
$\times$
($\lambda_\nu, \mu_\nu$)\}
$\rightarrow$
($\lambda, \mu$)
into account and numbering multiplicities in this
product by a running
integer index $\rho$.
The total orbital angular momentum $L$ is determined
via the chain $SU(3)
\supset SO(3)$ where $\kappa$ labels its multiplicity.
Likewise, the total spin of the protons, $S_\pi$, and
neutrons, $S_\nu$,
are coupled to the total spin of the nucleus using the
usual rules for
coupling angular momentum.
And finally, the total angular momentum $J$ results
from the coupling of
total orbital angular momentum $L$ with the total spin
$S$.

These group structures, reductions, and couplings are
an integral part of
the pseudo-SU(3) model and all of this information
enters
the corresponding wave functions:
\begin{eqnarray}
\lefteqn{
| \left\{ N_\pi [f_\pi] \alpha_\pi \, (\la_\pi,\,\mu_\pi) \, ,
          N_\nu [f_\nu] \alpha_\nu \, (\la_\nu,\,\mu_\nu) \,
\right\}
          \rho \, (\la,\,\mu) \kappa L \,
         \left\{S_\pi , S_\nu\right\} \, S \, ; JM \rangle
        }
\nn\\
&&\nn\\
&=&\sum_{\{ -\} }
\langle \,    (\la_\pi,\,\mu_\pi) \,  \kappa_\pi L_\pi
M_{L_\pi}\, ;
              (\la_\nu,\,\mu_\nu) \,  \kappa_\nu L_\nu
M_{L_\nu}\,
              | \,
              (\la,\,\mu) \,  \kappa  L M_L  \,
              \rangle_\rho
\nn \\
&&\nn\\
&&\times \, \langle \, S_\pi M_{S_\pi}, \,
                                                              S_\nu
M_{S_\nu}\,| \,
                                                              S  M_S
      \rangle  \, \times \,
\langle \, L M_L, \,
                  S M_S \,| \,
                   JM
\rangle  \, \nn\\
&&\nn\\
&& \times \,
         |N_\pi [f_\pi]   \alpha _\pi \,  (\la_\pi,\,\mu_\pi) \,
         \kappa_\pi L_\pi M_{L_\pi}\,  S_\pi M_{S_\pi}
         \rangle
\,\times \,
         |N_\nu [f_\nu]   \alpha _\nu \,
(\la_\nu,\,\mu_\nu) \,
         \kappa_\nu L_\nu M_{L_\nu}\,  S_\nu
M_{S_\nu}
         \rangle
\nn\\
&&\nn\\
&=&\sum_{\{ -\} }
\langle \, (\la_\pi,\,\mu_\pi) \,  \kappa_\pi L_\pi \, ;
           (\la_\nu,\,\mu_\nu) \,  \kappa_\nu L_\nu \,
           || \,
          (\la,\,\mu) \,  \kappa  L  \,
          \rangle_\rho
\, \times \,
\langle \, L_\pi M_{L_\pi}, \,
           L_\nu M_{L_\nu}\,| \,
           L M_L
\rangle  \,  \nn \\
&&\nn\\
&&\times \, \langle \, S_\pi M_{S_\pi}, \,
           S_\nu M_{S_\nu}\,| \,
           S  M_S
          \rangle  \,
\times \,
           \langle \, L M_L, \,
           S M_S \,| \,
           JM
           \rangle  \nn\\
&&\nn \\
&& \times \,
|N_\pi [f_\pi]   \alpha _\pi \,  (\la_\pi,\,\mu_\pi) \,
          \kappa_\pi L_\pi M_{L_\pi}\,  S_\pi M_{S_\pi}
          \rangle
\,
\times \,
|N_\nu [f_\nu]   \alpha _\nu \,  (\la_\nu,\,\mu_\nu) \,
         \kappa_\nu L_\nu M_{L_\nu}\,  S_\nu
M_{S_\nu}
         \rangle
\nn\\
\label{eq:wf}
\end{eqnarray}
with the abbreviation
$\{ - \}  = \{  M_{S_\pi}, \,  M_{S_\nu}, \,   M_{S},
\,
                      M_{L_\pi}, \,  M_{L_\nu}, \,   M_{L},
\,
                       \kappa_\pi, \, \kappa_\nu, \, L_\pi, \,
L_\nu
                 \} \, .
$
In this result the
$\langle \, L_1 M_{L_1}, \,
            L_2 M_{L_2}\,| \,
            L M_L
\rangle  $
factors are SU(2) Clebsch-Gordan coefficients and the
SU(3) coupling coefficients
$\langle \ldots;\, \ldots | \dots \rangle $
and
$\langle \ldots;\, \ldots || \dots \rangle $
are defined in \cite{AD73}.
The notation of Eq.(\ref{eq:wf}) indicates the
following couplings:
\begin{itemize}
\item SU(2):
$[L_\pi \times L_\nu ]^{LM_L}$,
$[S_\pi \times S_\nu ]^{SM_S }$, and
$[L \times S ]^{JM}$
where
$[A \times B ]^{CM_C}$ denotes the usual angular
momentum coupling \cite{EG87,VM88};
\item SU(3):
$\{ (\lambda_\pi,\, \mu_\pi) \times   (\lambda_\nu,\,
\mu_\nu) \} \rightarrow
\{ \rho \, (\lambda, \mu)\,  \kappa \, L \, M  \}
$ where the multiplicity index $\rho$ numbers the
multiple occurences of
$(\lambda, \mu)$ in the product
$\{ (\lambda_\pi,\, \mu_\pi) \times   (\lambda_\nu,\,
\mu_\nu) \}$.
\end{itemize}

\subsection{Operators and their matrix elements}
As explained in the introduction, accepting the fact that
SU(3) is a good
pseudo-space symmetry is a fundamental assumption
of the \psm.
The validity of this approximation, which applies to
nuclei with $A \ge 100$,
is based on:
\begin{itemize}
\item
the goodness of pseudo-spin symmetry, which is
suggested by the Nilsson
model and confirmed by other more sophisticated
theoretical analyses
\cite{RR73,TD94,BD92};
\item
the observation that most heavy nuclei are deformed,
which is important
because the goodness of the SU(3) symmetry increases
with increasing deformation.
\end{itemize}
And as shown in Section 2, there is a group-subgroup
chain for labeling
basis states that has SU(3) as an integral part of its
structure, and
which therefore takes full advantage of the \psma
symmetry.
To guarantee wavefunction-operator compatibility, it is
also necessary to
express operators associated with physical observables
in terms of SU(3)
tensors.
The procedure for doing this can be found in reference
\cite{CD87}.
In this contribution, the spin and proton-neutron
degrees of freedom
are explicitly incorporated into that formalism.
For the sake of completeness, it is useful to repeat
some of the material
given in reference \cite{CD87} as it forms a backdrop
for the introduction
of an extended notation that allows pairing correlations
to be incorporated
into the \psm.

The formalism requires one to know the relevant
group's coupling coefficients
and the tensorial properties of the single-particle
creation and annihilation
operators, which for a fermion system obey the usual
anticommutation relations:
\beq
\{ a^+_i, \, a_{k} \}_+ = \delta_{ik}, \, \, \,
\{ a^+_i, \, a^+_k \}_+ =  \{ a_i, \, a_k \}_+ = 0\, ,
\label{eq:comrel}
\eeq
where $i$ (and $k$) denote the full set of single-
particle quantum numbers
$\left\{ (\eta,0) \, lm \, \frac{1}{2}m_s \right\} $.
Here $(\eta, 0)$ are the single-particle SU(3) labels
\lmda and $l, \, \frac{1}{2},
$ $m$ and  $m_s $ stand, respectively, for the single-
particle orbital angular
momentum, spin, and their z-projections.
The effect of a creation operator on the single-particle
vacuum is given by
\beq
 a^+_{ (\eta,0) \, lm \, \frac{1}{2}m_s }
\, |0\rangle
=
| \,
(\eta,0) \, lm \, \frac{1}{2} m_s \rangle \, .
\eeq
While the creation operator $a^+_{ (\eta,0) \, lm \,
\frac{1}{2}m_s }$ is a
proper SU(3) tensor, the annihilation operator is not
and must be modified
appropriately to achieve this status
\cite{RR73,CD87,ET94}:
\beq
\tilde{a}_{(0,\eta) \, lm \ \frac{1}{2} m_s}
\equiv
(-1)^{\eta + l + m + \frac{1}{2} + m_s}
a_{(\eta,0) \, l-m \, \frac{1}{2} -m_s}\, .
\label{eq:an}
\eeq

With these elementary definitions in place, it is
possible to construct
three distinguishable tensor operators as products of
two creation and/or
annihilation operators:
\begin{itemize}
\item A one-body (unit tensor) operator,
\bea
^{(1,1)}{\cal F}^{\, \lm \,\kappa L M_L\, S
M_S}_{(\eta',0)(0,\eta) \frac{1}{2}
 \frac{1}{2}  }
 &\equiv&
\left\{
  a^+_{ (\eta',0) \, \frac{1}{2}  } \times  \tilde{a}_{
(0,\eta)  \, \frac{1}{2} }
\right\}^{\lm \,\kappa L M_L\,  S M_S }
\nn\\
&&\nn\\
&=&
\sum_{\{ - \} }
\langle \, (\eta', 0)\, 1 l' \, (0,\eta) \,  1 l ||
\lm \, \kappa L \rangle
\,
\times \langle \, l'  m' , \,
                   l m \,| \,
                   L M_L
\rangle
 \nn  \\
&&\nn\\
&&
\times
\langle \, \frac{1}{2} m^\prime_s , \,
                   \frac{1}{2} m_s \,| \,
                   S  M_S
\rangle
\, \, a^+_{ (\eta',0) \, l'm' \, \frac{1}{2}m^\prime_s }
\, \, \tilde{a}_{(0,\eta) \, lm \, \frac{1}{2} m_s} \,  ,
\label{eq:ut}
\eea
with the abbreviation
$
\{ - \}  = \{  l, \, m, \,  m_s, \,  l', \,  m', \, m^\prime_s
\}$ and where the left superscript denotes respectively
the number
of creation and annihilation operators in the product.
Note that the multiplicity index $\rho$ connected with
the
generic SU(3) product
$\{ (\la_1,\, \mu_1) \times   (\la_2,\, \mu_2) \}
\rightarrow
 \rho \,   (\lambda, \mu)
$
is usually dropped when the coupling is mutliplicity
free, that is,
whenever $\rho_{\mbox{max}}=1$.
To summarize, the couplings in Eq.(\ref{eq:ut}) are:
\begin{itemize}
\item SU(2):
$[l' \times l ]^{LM_L}$,
$[\frac{1}{2} \times \frac{1}{2}]^{SM_S}$\,;
\item SU(3):
$\{ (\eta',\, 0) \times   (0,\, \eta) \} \rightarrow
\{ \rho_{\mbox{max}}=1 \, ,  (\lambda, \mu) \}
$ \,.
\end{itemize}
\item An operator that creates a pair of particles,
\bea
^{(2,0)}{\cal F}^{\, \lm \,\kappa L M_L\, S
M_S}_{(\eta',0)(\eta, 0) \frac{1}{2}
 \frac{1}{2}  }
 &\equiv&
\left\{
  a^+_{ (\eta',0) \, \frac{1}{2}  } \times  a^+_{ (\eta,
0)  \, \frac{1}{2} }
\right\}^{\lm \,\kappa L M_L\,  S M_S }\, ,
\nn\\
&&\nn
\eea
where the coefficients that enter to effect the SU(2) and
SU(3) couplings in
this case are similar to those that appear in
Eq.(\ref{eq:ut}).
\item
A pair annihilation operator,
\bea
^{(0,2)}{\cal F}^{\, (\mu, \, \lambda) \,\kappa L
M_L\, S M_S}_{(0, \eta')(0,\eta) \frac{1}{2}
 \frac{1}{2}  }
 &\equiv&
\left\{
 \tilde{a}_{ (0, \eta') \, \frac{1}{2}  } \times
\tilde{a}_{ (0,\eta)  \, \frac{1}{2} }
\right\}^{(\mu, \, \lambda) \,\kappa L M_L\,  S M_S
}\, .
\nn
\eea
\end{itemize}

In a completely analogous manner, it is possible to
define two-body SU(3) unit
tensors as coupled products of the ${^{(2,0)}\cal F}$
and $^{(0,2)}{\cal F}$
operators:
\bea
^{(2,2)}{\cal F}^{\rho\, \lm \,\kappa L M_L\,
S M_S}_{(\la_1,\, \mu_1)(\mu_2,\, \la_2)\,  S_1\,
S_2  }
 &\equiv&
\left\{
^{(2,0)}{\cal F}^{ (\la_1, \, \mu_1)\, S_1
}_{(\eta_1',0)(\eta_1, 0) \frac{1}{2} \frac{1}{2}  }
\, \times \,
^{(0,2)}{\cal F}^{ (\mu_2, \, \la_2)\, S_2
}_{(0, \eta_2')(0,\eta_2) \frac{1}{2} \frac{1}{2}  }
\right\}^{\rho\, \lm \,\kappa L M_L\, S M_S}\, .
\eea
Note, that the indices
$ \eta_1',  \, \eta_1, \, \eta_2', \, \eta_2$
are suppressed in
$^{(2,2)}{\cal F}$.
This simplifies the notation and can be done because
these quantum numbers
enter into a \psma calculation as fixed parameters.
In fact, the constraints
$\eta_1' = \eta_1$, $\eta_2' = \eta_2$, and $\eta_1 =
\eta_2$ apply since
inter-shell couplings are not part of a $0\hbar\omega$
theory.
These  restrictions are lifted in a pseudo-symplectic
extension of the \psma
\cite{TD94}, and therefore for such applications the
notation must be expanded
appropriately.\footnote{A complete theory would also
include a general discussion
of non-particle number conserving as well as particle
number conserving operators
in both normal-ordered (creation operators to the left of
annihilation operators)
and multipole (for example, products of one-body
tensors) forms.
Indeed, it should be clear that an $n$-body operator
($^{(n,n)}{\cal F}$) can always
be expanded in multipole form (and vice-versa, of
course) by using the fundamental
commutation relations, Eq.(\ref{eq:comrel}), together
with recoupling formulae to
appropriately rearrange (group/order) the creation and
annihilation operators (see
\cite{TD94a}).
Non-particle number conserving operators
($^{(m,n)}{\cal F}$ with $m$ $\ne$ $n$),
represent pickup (stripping) type phenomena when
$m$ $<$ $n$ ($m$ $>$ $n$).
Here it is enough to limit the development to the forms
already introduced as
these suffice for our Hamiltonian, inclusive of the
pairing interaction.}

The definitions of
$^{(1,1)}{\cal F}$
and
$^{(2,2)}{\cal F}$
as one-body and two-body operators, respectively, are
applicable to identical
particle systems.
The next logical step is the introduction of proton-
neutron SU(3)-coupled
operators.
For example, the simplest multipole-multipole proton-
neutron tensor operator
can be defined as follows:
\bea
{\cal ^{\pi\nu}F}
^{\rho_o \,(\lambda_o,  \, \mu_o)\,  \kappa_o L_o
S_o;\, J_o M_o}
_{(\lambda_{\pi_o}, \, \mu_{\pi_o})
  (\lambda_{\nu_o}, \, \mu_{\nu_o}) \, S_{\pi_o} \,
S_{\nu_o}}
&\equiv&
\left\{  {\cal F}^{(\lambda_{\pi_o}, \, \mu_{\pi_o}) \,
S_{\pi_o}}
                 _{(\eta'_{\pi_o},0)(0,\eta_{\pi_o})
                   \frac{1}{2} \frac{1}{2}  }
\times
         {\cal F}^{(\lambda_{\nu_o},  \, \mu_{\nu_o}) \,
S_{\nu_o}}
                 _{(\eta'_{\nu_o},0)(0,\eta_{\nu_o})
                   \frac{1}{2} \frac{1}{2}  }
\right\}^{ \rho_o\, (\lambda_o,  \, \mu_o)\,  \kappa_o
L_o S_o ;\, J_o M_o}
\nn\\
\el
&=&
\sum_{M_{L_o}, M_{S_o}}
\langle  \,        L_o M_{L_o}, \,
                   S_o M_{S_o}  \,| \,
                   J_o M_o
\rangle
\nn\\
&&\nn\\
&&\times
\left\{ {\cal F}^{(\lambda_{\pi_o}, \, \mu_{\pi_o}) \,
S_{\pi_o}}
                _{(\eta'_{\pi_o},0)(0,\eta_{\pi_o})
\frac{1}{2} \frac{1}{2}}
\times
        {\cal F}^{(\lambda_{\nu_o},  \, \mu_{\nu_o}) \,
S_{\nu_o}}
                _{(\eta'_{\nu_o},0)(0,\eta_{\nu_o})
\frac{1}{2} \frac{1}{2}}
\right\} ^{\rho\,  (\lambda_o,  \, \mu_o)\,  \kappa_o
L_o M_{L_o} S_o M_{S_o}}.
\nn\\
&&\label{eq:gut}
\eea
The subscript ``$o$'' in this expression refers to
``operator'' and is introduced
in view of the forthcoming calculation of matrix
elements in the \psm.
The left superscript $\pi\nu$ on the ${^{\pi\nu}\cal
F}$ operator, which is a
shorthand notation for the complete label
\{$(m_{\pi},n_{\pi}) (m_{\nu},n_{\nu})$\} =
\{$(1,1) (1,1)$\},
denotes the fact that this operator acts simultaneously
in the proton and neutron
spaces.
The left superscript on each of the ${\cal F}$ factors in
Eq.(\ref{eq:gut}) is
suppressed because it is (1,1) in both cases, that is,
${\cal F}$ = $^{(1,1)}{\cal F}$.

In general a more explicit and complete notation is
required:
\bea
&&\nn\\
\lefteqn{
{^{\pi\nu}\cal F}
 ^{ \rho_o (\lambda_o,  \, \mu_o)\,  \kappa_o L_o
S_o;\, J_o M_o }
 _{ \{ \pi \} \, \{ \nu \} }
}
\nn\\
&=&
\left\{  {^{\pi}\cal F}^{\rho_{\pi_o}\,
(\lambda_{\pi_o}, \, \mu_{\pi_o}) \, S_{\pi_o}}
          _{\alpha_{\pi_1}(\la_{\pi_1}, \,
\mu_{\pi_1})\alpha_{\pi_2}(\mu_{\pi_2},
\, \la_{\pi_2})
            \, S_{\pi_1} S_{\pi_2}  }
\times
         {^{\nu}\cal F}^{\rho_{\nu_o}\,
(\lambda_{\nu_o}, \, \mu_{\nu_o}) \, S_{\nu_o}}
          _{\alpha_{\nu_1}(\la_{\nu_1}, \,
\mu_{\nu_1})\alpha_{\nu_2}(\mu_{\nu_2},
\, \la_{\nu_2})
            \, S_{\nu_1} S_{\nu_2}  }
\right\}^{ \rho_o\, (\lambda_o,  \, \mu_o)\,  \kappa_o
L_o S_o ;\, J_o M_o},
\nn\\
&&\label{eq:png}
\eea
where $\{ \pi \}$ stands for
$\left\{ \, \left[\alpha_{\pi_1}(\lambda_{\pi_1}, \,
\mu_{\pi_1}), \, S_{\pi_1} \right]
            \times
            \left[\alpha_{\pi_2}(\mu_{\pi_2}, \,
\la_{\pi_2}), \, S_{\pi_2} \right]
            \rightarrow
            \rho_{\pi_o} \, (\la_{\pi_o}, \, \mu_{\pi_o}),
\, S_{\pi_o}
\right\}
$
and similarly for $\{ \nu \}$, and the $\alpha_\pi$ and
$\alpha_\nu$ labels represent
additional quantum numbers that are required for a
unique identification of the
$^{\pi}\cal F$ and $^{\nu}\cal F$ factors in the
product.
In this case the left superscripts $\pi$, $\nu$, and
$\pi\nu$ stand for the sets
$(m_\pi,n_\pi)$, $(m_\nu,n_\nu)$, and
\{$(m_\pi,n_\pi)(m_\nu,n_\nu)$\}, respectively.
The matrix elements of such operators can be
calculated if the matrix elements of
the factors are known:
\bea
&&\nn\\
\lefteqn{ \langle
          \left\{ N'_\pi [f'_\pi]   \alpha' _\pi \,
(\la'_\pi,\,\mu'_\pi) \, ,
                  N'_\nu [f'_\nu] \alpha'_\nu \,
(\la'_\nu,\,\mu'_\nu) \, \right\}
               \, \rho' (\la', \,\mu') \,  \kappa' L' \,
          \left\{S'_\pi , S'_\nu\right\} \, S' ;\,  J' }
 \nn\\
&&\nn\\
&&
||
 {^{\pi\nu}\cal F}
  ^{\rho_o \,(\lambda_o,  \, \mu_o)\,  \kappa_o L_o
S_o;\, J_o}
  _{\{ \pi \} \, \{ \nu \}}
|| \nn \\
&&\nn\\
&&
\mbox{         }\left\{ N_\pi [f_\pi]   \alpha_\pi \,
(\la_\pi,\,\mu_\pi) \, ,
                        N_\nu [f_\nu] \alpha_\nu \,
                       (\la_\nu,\,\mu_\nu) \, \right\}
                    \,\rho\, (\la,\,\mu) \,
                          \kappa L \,
                  \left\{S_\pi , S_\nu\right\} \, S\, ; J \rangle
\nn\\
&& \nn\\
&&=
\chi
\left\{
\begin{array}{ccc}
                     L & L_o & L' \\
                     S  & S_o & S'  \\
                     J   & J_o  &  J'
\end{array}
\right\}
\,\,
\chi
\left\{
\begin{array}{ccc}
                    S_\pi  & S_{\pi_o} & S'_\pi \\
                    S_\nu  & S_{\nu_o} & S'_\nu \\
                    S & S_o & S'
\end{array}
\right\}
\sum_{ \bar{\rho} }
\langle \,
                 (\la,\,\mu) \,  \kappa L\, ,
                 (\la_o, \,\mu_o) \,  \kappa_o L_o \, || \,
                 (\la,'\,\mu') \,  \kappa'  L'  \,
\rangle_{\bar{\rho}}   \nn\\
&&\nn\\
&&\mbox{     } \times \sum_{\rho_\pi \rho_\nu}
\left\{
\begin{array}{cccc}
(\lambda_\pi, \,  \mu_\pi ) &
(\lambda_{\pi_o}, \,  \mu_{\pi_o} ) &
(\lambda'_\pi, \,  \mu'_\pi ) &
\rho_\pi \\
(\lambda_\nu, \,  \mu_\nu ) &
(\lambda_{\nu_o}, \,  \mu_{\nu_o }) &
(\lambda'_\nu, \,  \mu'_\nu ) &
\rho_\nu \\
(\lambda, \,  \mu ) &
(\lambda_o, \,  \mu_o ) &
(\lambda', \,  \mu' ) &
\bar{\rho} \\
\rho& \rho_o & \rho' & \\
\end{array}
\right\} \nn \\
&&\nn\\
&&\mbox{     } \times \langle N'_\pi [f'_\pi]
\alpha'_\pi
(\lambda'_\pi, \,  \mu'_\pi )\, S'_\pi \, |||
\,{^{\pi}\cal F}^{\rho_{\pi_o}\, (\lambda_{\pi_o}, \,
\mu_{\pi_o}) \, S_{\pi_o}}
                 _{\alpha_{\pi_1}(\la_{\pi_1}, \,
\mu_{\pi_1})\alpha_{\pi_2}(\mu_{\pi_2},
\, \la_{\pi_2})
                   \, S_{\pi_1} S_{\pi_2}  } \, ||| \,
N_\pi [f_\pi] \alpha_\pi (\lambda_\pi, \,  \mu_\pi )\,
S_\pi \rangle_{\rho_\pi} \nn\\
&&\nn\\
&& \mbox{     } \times \langle \, N'_\nu  [f'_\nu]
\alpha'_\nu
(\lambda'_\nu, \,  \mu'_\nu )\, S'_\nu \,\, |||\,\,
  {^{\nu}\cal F}^{\rho_{\nu_o}\, (\lambda_{\nu_o},
\, \mu_{\nu_o}) \, S_{\nu_o}}
                 _{\alpha_{\nu_1}(\la_{\nu_1}, \,
\mu_{\nu_1})\alpha_{\nu_2}(\mu_{\nu_2},
\, \la_{\nu_2})
                   \, S_{\nu_1} S_{\nu_2}  } \,\, ||| \,\,
N_\nu [f_\nu] \alpha_\nu (\lambda_\nu, \,  \mu_\nu )\,
S_\nu \, \rangle_{\rho_\nu }\, ,
\nn\\
\label{eq:ob}
\eea
where $\chi\{ \ldots \}$ denotes a unitary 9-j or Jahn-
Hope symbol \cite{VM88,JH54}
and
$\{ \dots \}$ its SU(3) extension \cite{Mi78}. The
triple-barred matrix elements that
enter in Eq.(\ref{eq:ob}),
$
\langle \ldots ||| \ldots ||| \ldots \rangle,
$
are reduced with respect to both SU(3) and SO(3) (see
Appendix \ref{appb}), and
can be evaluated for operators of physical interest with
the code introduced in
reference \cite{BD94}.
When this result is used to evaluate matrix elements of
either a pure proton
\{$(m_{\nu},n_{\nu}) = (0,0)$\} or a pure neutron
\{$(m_{\pi},n_{\pi}) = (0,0)$\}
operator, it simplifies just as for the corresponding
SU(2) case; namely, the
occurrence of a zero-body factor with its null SU(3)
irrep character means that
the 9-$(\lambda,\mu)$ coefficient can be reduced to a
6-$(\lambda,\mu)$ coefficient
and the corresponding triple-barred reduced matrix
element goes to unity.
The pairing interaction is a sum of two such operators:
$H_{P}$ $\rightarrow$ \{$^{(2,2)(0,0)}\cal F$ +
$^{(0,0)(2,2)}\cal F$\}.
\section{The pairing interaction}
As stated in the introduction, the inclusion of a short-
range interaction
of the pairing type in pseudo-SU(3) calculations has
not been tested and
seems from other independent evidence to be
important.
But before exploring the influence pairing has on
collective nuclear
properties, it is necessary to introduce some additional
formalism.
In second quantization the pairing Hamiltonian
\cite{BS69} is defined as:
\beq
H_P =
-\frac{G}{4} \sum_{\{-\}}
a^+_{\eta_1 l_1 \frac{1}{2}j_1m_1}
a^+_{\overline \eta_1 \overline l_1
\overline{\frac{1}{2}} \overline j_1 \overline m_1}
a_{\overline \eta_2 \overline l_2
\overline{\frac{1}{2}} \overline j_2 \overline m_2  }
          a_{\eta_2 l_2\frac{1}{2}j_2m_2}\, ,
\label{eq:hp1}
\eeq
where
$\{ - \} = \{ \eta_1 \, , l_1  \, ,  j_1 \, ,  m_1 \, ,
\eta_2 \, ,  l_2 \, , j_2 \, , m_2 \}$ with
$(m_i = -j_i, -j_i+1, \ldots, j_i)$ and $(i=1,2)$.
The bar over quantum numbers in Eq.(\ref{eq:hp1})
denotes time-reversal
and $G$ is the pairing strength, which is normally
taken to be somewhat
different for protons and neutrons.

Obviously $H_P$, as given in Eq.(\ref{eq:hp1}), is
defined in terms of
normal-space quantum numbers.
However, since all calculations in the \psma are
performed in the pseudo-space,
it is necessary to transform $H_P$ into its pseudo-
space representation.
An application of the normal $\rightarrow$ pseudo
mapping is described
in reference \cite{CD87} for the general case, and in
reference \cite{BE94} for
the special case of pairing.
The following is the final identical-particle result,
expressed in terms
of pseudo-space quantum numbers \cite{CB94}:
\bea
\lefteqn{H_P
= -\frac{G}{2}
\sum_{ \{ - \} }
(-1)^{l_1 + l_2}
\left( (2l_1 + 1) (2l_2+1)
\right)
^\frac{1}{2} \,
\langle \, (\eta_1, \, 0) \, 1 \, l_1; \,
           (\eta_1, \, 0) \, 1 \, l_1
           ||
           (\la_1, \,  \mu_1 )\,  1 \,  0
\rangle_{1}
}\nn\\
\el
&\times&
\sutrcc{\sutrqn{\eta_2}{0}\, 1\, {l_2}}
       {\sutrqn{\eta_2}{0}\, 1\, {l_2}}
       {\sutrqn{\la_2 }{\mu_2 }\, 1\, {0}}_{1}
\,
\sutrcc{\sutrqn{\la_1}{\mu_1}\, 1 \, {0}}
       {\sutrqn{\mu_2}{\la_2}\, 1 \, {0}}
       {\sutrqn{\la  }{\mu }\, 1  \, {0}}_\rho \nn\\
\el
&\times &
{^{(2,2)}\cal F}^{\rho\, \lm \,\kappa=1 \, L=0 \, S=0
\, J=0 \,
}_{(\la_1,\, \mu_1)(\mu_2,\, \la_2)\,  S_1=0\, S_2=0
}
\nn\\
&&
\label{eq:pair}
\eea
where
$\{ -\} =
\{l_1, l_2, \la_1, \mu_1, \la_2, \mu_2, \rho, \la, \mu \}
$.
(The summation over $\eta_1$ and $\eta_2$ is dropped
because
the pseudo-SU(3) model is a $0\hbar \omega$ theory
and inter-shell couplings are not allowed.)
Since the pairing interaction acts only between like
nucleons, the proton (neutron)
pairing operator $H_P^\pi$ ($H_P^\nu$ ) can be
obtained by replacing the quantum
numbers in the generic expression, Eq.(\ref{eq:pair}),
by their proton (neutron)
counterparts.
And as already noted, since the proton (neutron)
pairing interaction is a SU(3)
scalar operator in the neutron (proton) subspace,
$(\lambda_{\nu}, \, \mu_{\nu}) = (0,0)$
($(\lambda_{\pi}, \, \mu_{\pi}) = (0,0)$),
the expression for its matrix elements simplify
considerably as compared to those
of a general proton-neutron interaction, see
Eq.(\ref{eq:ob}).
Results for the matrix elements of $H_P^\pi$ and
$H_P^\nu$ in a proton-neutron basis
are given in Appendix \ref{pn:pair}.
\section{Pairing and collective states}
Since the primary objective of this study is to assess
the importance of
pairing correlations in heavy nuclei as revealed within
the context of \psm,
it is of interest to study the influence of the pairing
interaction on the
wavefunctions themselves.
Consider the normal-space Hamiltonian
\beq
H_{QP} = \frac{\chi}{2} \,\,  {^rQ^a} \cdot \,
{^rQ^a} + G_\pi H^\pi_P + G_\nu H_P^\nu
+ a K_J^2 + bJ^2
\label{eq:spair}
\eeq
where $^rQ^a$ denotes the real quadrupole operator,
$J$ the collective
angular momentum operator, and  $K_J^2$ stands for
a residual interaction
that is designed to generate $K$-band splitting (see
\cite{ND92,NB94}).
Although the deformation driving ${^rQ^a} \cdot \,
{^rQ^a}$ term is not
exactly equal to its pseudo-space image $Q^a \cdot
Q^a$, the difference
is small.
Indeed, within the leading SU(3) irrep the induced
corrections to energies
and electromagnetic transition probabilities have been
shown to be smaller
than $1\%$ \cite{CL93}.
Since the exact results for the normal-space
$\rightarrow$ pseudo-space transformation are known for
$
H^\pi_P
$
and
$
H_P^\nu
$
(see Eqs. \ref{eq:ppair}, \ref{eq:npair}), and the
$K_J^2$ and $J^2$
operators transform as scalars,
\beq
H_{QP} = \frac{\chi}{2} \,\,  Q^a \cdot Q^a + G_\pi
H^\pi_P + G_\nu H_P^\nu
+ a K_J^2 + bJ^2
\label{eq:psh}
\eeq
for practical purposes.
(If not explicitly indicated otherwise, the parameters
have the numerical
values $\chi= 4.32$ keV, $a = 0.202$ MeV, and
$b=9.26$ keV which are compatible
with a realistic set taken from best-fit calculations for
heavy deformed
nuclei \cite{TB94}).

The results presented next refer to the relatively simple
system of two protons
and two neutrons in pseudo-shells $\eta_\pi = 3$ and
$\eta_\nu = 4$, respectively.
Systems with a larger number of particles have also
been studied and were found
to display the same qualitative behaviour as the $2\pi
2\nu$ system.
The results for these more complicated cases will be
published in \cite{TB94}
which focuses on complementary characteristics of the
pairing interaction.
An advantage of the present study is that concerns
associated with truncating the
model space can be avoided altogther because the $2\pi
2\nu$ system, though rich in
structure, is small enough to be solved without
invoking truncation measures.

Before taking a closer look at the quadrupole
$\leftrightarrow$ pairing strength
relation, it is instructive to consider the spectrum for
the case of pure pairing,
that is, when $\chi=a=b=0$ and $G_\pi = G_\nu = 1$
MeV in $H_{QP}$.
The results shown in Fig. \ref{fig:pupair},
(A), provide a check on the calculations since in this
case analytic results for
the excitation energies can be given, and,
(B), illustrate the structure of the pairing spectrum for a
proton-neutron
system.\\[1 cm]
{\bf Figure \ref{fig:pupair}}
\\[1 cm]
Analytical results for the pairing spectrum can be
derived
in a quasi-spin formalism \cite{RS80,CB94} with
identical particle states classified
according to their seniority $s$, which is simply the
number of particles not
coupled to J=0 pairs.
For the combined proton-neutron system, the
excitation energy $E_{s_\pi s_\nu}$ is
given by the sum
\[
E_{s_\pi s_\nu} = -\frac{G}{4} \sum_{\sigma \, =
\,\pi, \, \nu}
\left( s_{\sigma} [s_{\sigma} -
2(\Omega_{\sigma}+1)]
+2N_{\sigma}(\Omega_{\sigma}+1) - N_{\sigma}^2
\right)\, ,
\]
where $s_\sigma$ denotes the seniority, $N_\sigma$
the number of nucleons, and
$\Omega_\sigma$ the number of levels in the
$\eta_\sigma$ shell
($\pi \rightarrow$ proton,
$\nu \rightarrow$ neutron).
(Recall that $\Omega = (\eta + 1)(\eta + 2)/2)$ for the
$\eta$-th major (normal or pseudo) shell.)
Taking $N_\pi = N_\nu = 2$ with $\eta_\pi = 3$ and
$\eta_\nu = 4$ one
obtains a pairing gap of 10 MeV (15 MeV) for the
energy required to
break a proton (neutron) pair.
If one renormalizes the spectrum with respect to the
ground state one
obtains the results shown in Fig. \ref{fig:pupair}.
The states which are ordered according to increasing
angular momentum
group into degenerate sets classified by the seniority
quantum numbers
$(s_\pi, \, s_\nu)$.
The dimension of each set, which is given on the top
of each horizontal
bar,indicates that only a few states contribute to the
low energy spectrum.
The result of combining this pure pairing seniority
level scheme with a
quadrupole dominated rotational spectrum to yield
more realistic results
is considered next.

To investigate the role pairing plays in a description of
collective
phenomena, consider the $N_\pi = N_\nu = 2$ system
introduced above
but now with the pairing and quadrupole interaction
strengths both
taken to be non-zero.
Specifically, the quadrupole strength $\chi$ was fixed
at a realistic
value and the pairing strength was varied from 0 to
0.30 MeV.
(For simplicity, the proton and neutron pairing
strengths were set
equal to one another, $G_\pi = G_\nu \equiv G$.)
An estimate for realistic $G_\pi$ and $G_\nu$ values
can be
obtained from various phenomenological formulas, for
example, in
\cite{Ca90} one finds the result $G_\pi =
\frac{17}{A}$ MeV and
$G_\nu = \frac{23}{A}$ MeV.
For our example of a rare earth nucleus with 2 protons
(neutrons) in
the pseudo shell $\eta_\pi=3$ ($\eta_\nu=4$), which
derives from a
normal shell with principal quantum number
$\eta_\pi=4$ ($\eta_\nu=5$),
one has $A=136$ and hence $G_\pi = 0.125$ MeV
($G_\nu = 0.169$ MeV).
(Although this system happens to have the proton and
neutron numbers of
$^{136}_{52}\mbox{Xe}_{84}$, it is important to
remember that the purpose
of this investigation is to study the influence of the
pairing interaction
on excitation spectra and not to model a particular
physical system.
This exercise should be considered a forerunner for
forthcoming attempts
at describing and predicting experimental data.
Best fit calculations for
$^{136}_{52}\mbox{Xe}_{84}$ as well as
other rare-earth nuclei will be published elsewhere
\cite{TB94}.)
\\[1 cm]
{\bf Figure \ref{fig:l0}, \ref{fig:l4}, \ref{fig:l8} }
\\[1 cm]
Figures \ref{fig:l0}, \ref{fig:l4}, and \ref{fig:l8}
indicate absolute
values of the amplitudes (upper plots) and intensities
(lower plots)
of the calculated $0_1$, $4_1$, and $8_1$ eigenstates.
These values
(vertical axis) are plotted as a function of the basis state
number
(running from left to right) and the pairing strength
(increasing from
front to rear).
The basis states have been sorted as a function of the
eigenvalue of
the second order SU(3) Casimir invariant, $C_2=\la^2
+ \la\mu + \mu^2
+ 3(\la + \mu)$.
Hence, in each figure the most deformed SU(3) irrep is
associated
with the index 1, which is on the far left, and the least
deformed
with the highest index on the far right.
The pairing strength is indicated on the right, where for
simplicity
the strengths were set equal to one another, $G_\pi =
G_\nu \equiv G$.

For $G=0$ the quadrupole interaction dominates and
the ground state
wavefunction is comprised of the most deformed
SU(3) irrep only; that
is, no mixing occurs, and this result is angular
momentum independent.
As the pairing strength is increased, contributions from
irreps that
correspond to less deformed configurations grow, but
not all basis
states contribute in the same way to each eigenstate.
Instead, the results suggest that there may be a
distinguishable
pattern that applies for each value of the total angular
momentum.
If this turns out to be the case and can be predicted
apriori, it
could lead to a prescription for an angular momentum
dependent
model space truncation.
This subject is currently under investigation.
Regardless of whether this is true or not, the point to
note is that
for realisitic values of the pairing strength
$(0.1$ MeV $\leq G \leq$ $0.2$ MeV)
there is significant basis state mixing for all values of
the total
angular momenta --- the pairing interaction definitely
breaks the
SU(3) symmetry.

Notice that the contribution of the leading irrep
increases with
increasing angular momentum $L$.
Two complementary factors contribute to this effect:
First of all, within any $U(\Omega)$ representation the
number of
SU(3) irreps that contribute to a particular $L$ value
decreases
as $L$ increases.
Indeed, if $(\lambda,\mu)$ is the leading SU(3) irrep
then the value
$L = (\lambda + \mu)$ is unique and the probability of
it occuring in
the eigenstate with $J = L = (\lambda + \mu)$ is unity
$(100\%)$,
independent of the nature of the interaction.
Secondly, as the value of $L$ increases, the
probability of finding
pairs that couple to angular momentum zero decreases
because the
formation of a zero-coupled pair subtracts angular
momentum from
the system.
Another way of saying the same thing is that the
formation of a
pair effectively reduces the system to one with two less
particles
and unless one is near the mid-shell region, the leading
SU(3) irrep
of such a system has a lower maximum $L$ value.

The influence of the pairing strength $G$ on the
rotational ground
state band of the $N_\pi = N_\nu = 2$ example is
shown in
Fig. \ref{fig:gsb}.
\\[1 cm]
{\bf Figure \ref{fig:gsb}}
\\[1 cm]
With the same parameters as used in the above
analyses, one clearly
finds that with $G$ increasing from
$0.0$MeV to $0.3$MeV the system moves from a
pure rotational structure
towards one that is like that of the pure pairing
Hamiltonian,
see Fig. \ref{fig:pupair} for comparison.
For realistic values of $G$ $(0.1$ MeV $\leq G \leq$
$0.2$ MeV)
there is clearly competition between the quadrupole
dominated
rotational picture and the pairing dominated seniority
scheme.

Effective moments of inertia $\theta$ for each of the
excited levels
in Fig. \ref{fig:gsb} are shown in Fig. \ref{fig:mi}.
The values plotted were extracted from the spectrum
using the simple
formula $\theta = \frac{L(L+1)}{2E}$.
The fact that the moments of inertia decrease with
increasing pairing
strength is to be expected because typical $2^+$ states
of rotational
bands lie at a few tens of keV while those of pairing-
dominated
vibrator configurations are usually in the MeV range.
Note that for a fixed value of the pairing strength the
magnitude
of the decrease in the moments of inertia is less the
higher the
$L$ value.
Again, this is the expected result because states with
high angular
momentum as compared to those with low values resist
the formation
of pairs because each pair subtracts angular momentum
from the system,
thereby making rotation more difficult to generate.
So while from Fig. \ref{fig:gsb} an increase in the
pairing appears
to affect states with high $L$ more strongly than those
with lower
$L$ values, the opposite holds. The former is an
illusion created by
the $L(L+1)$ multiplier that enters into the energy
equation; the
pairing interaction is most effective in states with low
$L$ values.
This picture is consistent, for example, with the
interpretation of
backbending as being associated with the break-up of a
pair.

\section{Summary and outlook}
The formalism that is required for carrying out shell-
model calculations
within the framework of the \psm, extended to include
explicitly both the
spin and proton-neutron degrees of freedom, has been
introduced.
In particular, the pairing interaction, which couples
different irreps of
SU(3), has been expressed in terms of pseudo-space
unit tensor operators,
and general expressions for calculating the matrix
elements of these tensors,
and hence the pairing interaction have been
given.
(Another study will give a more detailed account of the
pairing force and explore additional features
that reflect on the importance of pairing correlations in
heavy nuclei
\cite{CB94}.)
The presented results make liberal use of an extended
technology for calculating
SU(3) coupling and recoupling coefficients
\cite{AD73} and a recently
released code for evaluating SU(3) reduced matrix
elements \cite{BD94}.

The influence of pairing correlations on the collective
(rotational) properties
of the $N_\pi = 2$ ($\eta_\pi = 3$) and $N_\nu = 2$
($\eta_\nu = 4$) system was
then considered.
The Hamiltonian that was used includes a $Q^a \cdot
Q^a$ term, the pairing
interaction, and residual $J^2$ and $K_J^2$ terms that
serve to generate the
$J(J+1)$ and $K$-band splitting.
(The splitting generated by $J^2$ is trivial because
rotational invariance
insures that $J$ is a good quantum number, but since
$K_{J}$ is not an exact
symmetry the splitting generated by $K_J^2$ is non-
trivial, and can compete,
for example, with that generated by the pairing
interaction.)
For vanishing pairing strengths ($G_\pi = G_\nu =
0$), this Hamiltonian has
non-vanishing matrix elements between states of the
same irrep only.
In particular, for $J=0$ states neither the $J^2$ and
$K_J^2$ terms contribute
so only the $Q^a \cdot Q^a$ term enters and as a
consequence the leading SU(3)
irrep, which is unique, is the only contributor to the
ground state.
Upon increasing the pairing strength, however, this
perfect SU(3) symmetry is
broken and other irreps enter, especially in the ground
state.
Similar results are observed for the states of higher
angular momentum, but
to a lesser degree because for these the role of the
pairing interaction is
curtailed and the $K_J^2$ term enters.

This paper is the first in a series of contributions which
are aimed at
achieving a proper quantitative description of heavy
nuclei within the
framework of the \psm, with pairing playing its proper
role.
In keeping with this plan, the general formalism for
calculating the matrix
elements of any interaction was introduced, with
explicit results given for
the pairing interaction.
To gain a deeper understanding of the importance of
short-range correlations
in strongly deformed systems, it will certainly be
useful to investigate the
effect pairing has on other collective nuclear properties
such as $g_R$ factors,
B(M1) values, $K_J$ mixing, B(E2) values, etc.
Such studies will serve as the basis for a quantitative
description of heavy
nuclei within the framework of the \psm.
Beyond these studies, however, lies an even bigger
challenge, namely, the development
of a comprehensive program aimed at the unification of
shell-model and collective-model
theories.
It is our goal to be able to give a proper account of
macroscopic phenomena in terms
of a user-friendly and usable microscopic theory.
The simplifications provided by the pseudo-spin
concept are, of course, key to
achieving this lofty objective.

\newpage
\section{Appendix}
\appendix
\section{Proton and neutron pairing matrix elements
\label{pn:pair}}
Applying the general result given in Section 2 for
SU(3) matrix elements
to the proton pairing operator, see Eq.(\ref{eq:pair}),
yields:
\bea
\lefteqn{  \langle
                  \left\{ N'_\pi [f'_\pi]   \alpha' _\pi \,
(\la'_\pi,\,\mu'_\pi) \, ,
                            N'_\nu [f'_\nu] \alpha'_\nu \,
(\la'_\nu,\,\mu'_\nu) \, \right\}
                            \, \rho' (\la', \,\mu') \,  \kappa' L' \,
                  \left\{S'_\pi , S'_\nu\right\} \, S' ;\,  J' }
 \nn\\
\el
&&
||
H_P^{\pi}
|| \nn \\
\el
&&
\mbox{         }\left\{ N_\pi [f_\pi]   \alpha_\pi \,
(\la_\pi,\,\mu_\pi) \, ,                            N_\nu [f_\nu]
\alpha_\nu \,
                           (\la_\nu,\,\mu_\nu) \, \right\}
                           \, \rho  \, (\la,\,\mu) \,
                          \kappa L \,
                  \left\{S_\pi , S_\nu\right\} \, S\, ; J \rangle
\nn\\
&& \nn\\
&&=
\delta_{LL'} \delta_{SS'} \delta_{JJ'} \delta_{S_\pi
S_\pi'} \delta_{S_\nu S_\nu'}
\delta_{\al_{\nu}\al_{\nu'}}
\delta_{\la_{\nu}\la_{\nu'}}
\delta_{\mu_{\nu}\mu_{\nu'}}
\sqrt{ \frac{2J+1}{(2L+1) (2S_\pi+1)(2S_\nu+1)}}
\nn\\
\el
&&\times
\sum_{\la_0, \, \mu_0, \bar{\rho}}
\langle \,
                 (\la,\,\mu) \,  \kappa L\, ,
                 (\la_o, \,\mu_o) \,  \kappa_o=1 L_o=0 \, || \,
                 (\la',\,\mu') \,  \kappa'  L'  \,
\rangle_{\bar{\rho}}
\nn\\
&&\nn\\
&& \times \sum_{\rho_\pi}
\left\{
\begin{array}{cccc}
(\lambda_\pi, \,  \mu_\pi ) &
(\lambda_o, \,  \mu_o ) &
(\lambda'_\pi, \,  \mu'_\pi ) &
\rho_\pi \\
(\lambda_\nu, \,  \mu_\nu ) &
 (0, \,  0 ) &
(\lambda'_\nu, \,  \mu'_\nu ) &
 1 \\
(\lambda, \,  \mu ) &
(\lambda_o, \,  \mu_o ) &
(\lambda', \,  \mu' ) &
\bar{\rho} \\
\rho& 1 & \rho' & \\
\end{array}
\right\}
\nn\\
\el
&&\times
\sum_{ \{ - \} }
C^{\, \rho_{12} \, (\lambda_o,  \, \mu_o)}_{\, (\la_1,
\, \mu_1) \, (\mu_2, \, \la_2) }
\langle N'_\pi [f'_\pi] \alpha'_\pi
(\lambda'_\pi, \,  \mu'_\pi ), S'_\pi \,
|||
{\cal T}^{\rho_{12} \, (\lambda_o,  \, \mu_o)
\,S_{\pi_o}=0
}_{\, (\la_1, \, \mu_1) \, (\mu_2, \, \la_2) } \,
|||
 \,
      N_\pi [f_\pi] \alpha_\pi (\lambda_\pi, \,  \mu_\pi ),
S_\pi
\rangle_{\rho_\pi} \nn\\
&&\label{eq:ppair}
\eea
where
$\{ - \} = \{\la_1, \, \mu_1, \, \la_2, \, \mu_2, \,
\rho_{12} \} $.
The triple barred reduced matrix element that enters
here is defined
in Appendix \ref{appb}.
The expansion coefficients
$
C^{\, \rho_{12} \, (\lambda_o,  \, \mu_o)}_{\, (\la_1,
\, \mu_1) \, (\mu_2, \, \la_2) }
$
are given by:
\bea
C^{\, \rho_{12} \, (\lambda_o,  \, \mu_o)}_{\, (\la_1,
\, \mu_1) \, (\la_2, \, \mu_2) }
&\equiv&
\sum_{ l_1 l_2 }
\frac{G}{2}
(-1)^{l_1 + l_2 +1}
\left( (2l_1 + 1) (2l_2+1)
\right)^\frac{1}{2}
\nn\\
\el
&\times&
\langle \, (\eta_\pi, \, 0) \, 1 \, l_1; \,
           (\eta_\pi, \, 0) \, 1 \, l_1
           || \,
           (\la_1, \,  \mu_1 )\,  1 \,  L_1=0
\rangle_{1}
\nn\\
\el
&\times&
\sutrcc{\sutrqn{\eta_\pi}{0}\, 1\, {l_2}}
       {\sutrqn{\eta_\pi}{0}\, 1\, {l_2}}
       {\sutrqn{\la_2 }{\mu_2 }\, 1\, {L_2=0}}_{1}
\nn\\
\el
&\times&
\sutrcc{\sutrqn{\la_1}{\mu_1}\, 1 \, 0}
       {\sutrqn{\mu_2}{\la_2}\, 1 \, 0}
       {\sutrqn{\la  }{\mu }\, 1  \, 0}_{\rho_{12}}
\label{eq:cfactor}
\eea
where the constraint $\eta_{\pi_1} = \eta_{\pi_2}
\equiv \eta_\pi$ has been
invoked because within a \psma framework the action
of the pairing operator
is restricted to one major shell.  (See Chapter 7 of
\cite{CB94} for a more
complete story, including, among other things,
numerical results for the
expansion coefficients for the $\eta = 3$ case.)

A complementary result ($\pi \leftrightarrow \nu$, etc.)
can be used to calculate
matrix elements of the neutron pairing operator:
\bea
\lefteqn{  \langle
                  \left\{ N'_\pi [f'_\pi]   \alpha' _\pi \,
(\la'_\pi,\,\mu'_\pi) \, ,
                            N'_\nu [f'_\nu] \alpha'_\nu \,
(\la'_\nu,\,\mu'_\nu) \, \right\}
                            \, \rho' (\la', \,\mu') \,  \kappa' L' \,
                  \left\{S'_\pi , S'_\nu\right\} \, S' ;\,  J' }
 \nn\\
\el
&&
||
H_P^{\nu}
|| \nn \\
\el
&&
\mbox{         }\left\{ N_\pi [f_\pi]   \alpha_\pi \,
(\la_\pi,\,\mu_\pi) \, ,
                            N_\nu [f_\nu] \alpha_\nu \,
                           (\la_\nu,\,\mu_\nu) \, \right\}
                           \, \rho  \, (\la,\,\mu) \,
                          \kappa L \,
                  \left\{S_\pi , S_\nu\right\} \, S\, ; J \rangle
\nn\\
&& \nn\\
&&=
\delta_{LL'} \delta_{SS'} \delta_{JJ'} \delta_{S_\pi
S_\pi'} \delta_{S_\nu S_\nu'}
\delta_{\al_{\pi}\al_{\pi'}}
\delta_{\la_{\pi}\la_{\pi'}}
\delta_{\mu_{\pi}\mu_{\pi'}}
\sqrt{ \frac{2J+1}{(2L+1) (2S_\pi+1)(2S_\nu+1)}}
\nn\\
\el
&&\times
\sum_{\la_0, \, \mu_0, \bar{\rho}}
\langle \,
                 (\la,\,\mu) \,  \kappa L\, ,
                 (\la_o, \,\mu_o) \,  \kappa_o=1 L_o=0 \, || \,
                 (\la',\,\mu') \,  \kappa'  L'  \,
\rangle_{\bar{\rho}}
\nn\\
&&\nn\\
&& \times \sum_{\rho_\pi}
\left\{
\begin{array}{cccc}
(\lambda_\pi, \,  \mu_\pi ) &
( 0 , \,  0  ) &
(\lambda'_\pi, \,  \mu'_\pi ) &
1 \\
(\lambda_\nu, \,  \mu_\nu ) &
 (\lambda_o, \,  \mu_o ) &
(\lambda'_\nu, \,  \mu'_\nu ) &
 \rho_\nu \\
(\lambda, \,  \mu ) &
(\lambda_o, \,  \mu_o ) &
(\lambda', \,  \mu' ) &
\bar{\rho} \\
\rho& 1 & \rho' & \\
\end{array}
\right\}
\nn\\
\el
&&\times
\sum_{ \{ - \} }
C^{\, \rho_{12} \, (\lambda_o,  \, \mu_o)}_{\, (\la_1,
\, \mu_1) \, (\mu_2, \, \la_2) }
\langle N'_\nu [f'_\nu] \alpha'_\nu
(\lambda'_\nu, \,  \mu'_\nu ) S'_\nu \,
|||
{\cal T}^{\rho_{12} \, (\lambda_o,  \, \mu_o)
\,S_{\nu_o}=0
}_{\, (\la_1, \, \mu_1) \, (\mu_2, \, \la_2) } \,
|||
 \,
      N_\nu [f_\nu] \alpha_\nu (\lambda_\nu, \,  \mu_\nu
) S_\nu
\rangle_{\rho_\nu} \nn\\
&&\label{eq:npair}
\eea
with
$\{ - \} = \{\la_1, \, \mu_1, \, \la_2, \, \mu_2, \,
\rho_{12} \} $.
The
$
C^{\, \rho_{12} \, (\lambda_o,  \, \mu_o)}_{\, (\la_1,
\, \mu_1) \, (\mu_2, \, \la_2) }
$
in this case is for neutrons, that is, $\eta_\pi
\rightarrow \eta_\nu$ in
Eq.(\ref{eq:cfactor}).

\newpage
\section{Tripple-barred reduced matrix elements
\label{appb}}
The Wigner-Eckart theorem for SU(2) gives
\begin{eqnarray}
\lefteqn{
\langle \, (\la_3,\,\mu_3) \,  \kappa_3 l_3 m_3
|\,  T^{(\la_2,\,\mu_2) \kappa_2 l_2  m_2}\,
| \, (\la_1,\,\mu_1) \, \kappa_1 l_1 m_1\,\rangle
              }
\nn\\
&&\nn\\
&=&
\mbox{   }
\langle \, l_1 m_1,\, l_2 m_2 \, | \, l_3 m_3 \,\rangle
\,
\langle \, (\la_3,\,\mu_3) \, \kappa_3 l_3
||  \, T^{(\la_2,\,\mu_2)\, \kappa_2 l_2  } \,
|| \, (\la_1,\,\mu_1) \, \kappa_1 l_1\, , \rangle
\end{eqnarray}
where
$
\langle \, (\la_3,\,\mu_3) \,  \kappa_3 l_3
||  \, T^{(\la_2,\,\mu_2) \kappa_2 l_2}\,
|| \, (\la_1,\,\mu_1) \,  \kappa_1 l_1\,\rangle
$
is a double-barred SU(2)-reduced matrix element.
Analogously, it is possible to extend this factorization
to SU(3) by introducing
SU(3)-reduced triple-barred matrix elements. An
important difference in this case,
however, is a sum over the multiplicity index $\rho$:
\begin{eqnarray}
\lefteqn{
\langle \, (\la_3,\, \mu_3) \,  \kappa_3 l_3 m_3\,
|\,T^{(\la_2,\,\mu_2) \kappa_2 l_2  m_2}\, | \,
(\la_1,\,\mu_1) \, \kappa_1 l_1 m_1\,\rangle }
\nn\\
\el
&& \! \! \! \! \! \! = \!  \! \!  \! \,\,\,\,
\sum_\rho
\mbox{   }
\langle \, (\lao,\muo)\, \ko l_1 m_1
 ; \,
(\lat,  \mut) \, \kt l_2 m_2
|
\, (\la\mu) \, \k_3 l_3 m_3\rangle_\rho
\,
\langle \, (\la_3,\,\mu_3) \,
|||  \, T^{(\la_2,\,\mu_2)}\,
|| | \, (\la_1,\,\mu_1) \,\rangle_\rho \nn \\
\nn\\
&& \! \! \! \! \! \! = \! \! \!  \! \,\,\,\,
\sum_\rho
\mbox{   }
\langle \, l_1 m_1,\, l_2 m_2 \, | \, l_3 m_3 \,\rangle
\langle \, (\lao,\muo)\, \ko l_1
 ; \,
(\lat,  \mut) \, \kt l_2
||
\, (\la_3\mu_3) \, \k l_3\rangle_\rho
\,
\nn\\
\el
&& \, \, \, \, \,
\times
\langle \, (\la_3,\,\mu_3) \,
|||  \, T^{(\la_2,\,\mu_2)}\,
||| \, (\la_1,\,\mu_1) \,\rangle_\rho  \,  .\nn
\end{eqnarray}

\clearpage

\begin{center}
{\Large Figure captions}
\end{center}
\noindent
\begin{figure}[htb]
\caption{\label{fig:wf}
The group chains that are used for the classification of
pseudo-SU(3)
wavefunctions are depicted in the two center columns.
The eigenvalues of the corresponding Casimir
invariants, which are
indicated on the far left and right of the diagram,
provide a complete
labeling scheme for pseudo-SU(3) basis states.
}
\end{figure}
\noindent
\begin{figure}[htb]
\caption{\label{fig:pupair}
The excitation spectrum of $H_{QP}$ for the pseudo-
space configuration
$[\,(fp)^{N_\pi=2} (gds)^{N_\nu=2}\,]$ for the case
of a pure pairing
interaction, $\chi=0$ and $G_\pi = G_\nu = 1$ MeV.
The horizontal axis denotes the total angular
momentum with $L=J$ since
$S=0$.
The seniority quantum numbers, ($s_\pi$, $s_\nu$)
for protons and neutrons,
respectively, are indicated on the far right and the
numbers on top of each
level bar denote the corresponding dimension.
}
\end{figure}
\noindent
\begin{figure}[htb]
\caption{\label{fig:l0}
Spreading of the $L=0$ ground state over the 29
dimensional $L=0$ \psma basis
of the $[\,(fp)^{N_\pi=2} (gds)^{N_\nu=2}\,]$
system.
Absolute values of amplitudes (upper graphs) and
intensities (lower graphs)
are plotted on the vertical axis as a function of the basis
state number
(sorted according to the eigenvalue of the second order
Casimir invariant of
SU(3), $C_2=\la^2 + \la\mu + \mu^2  + 3(\la + \mu)$,
and therefore according
to their intrinsic deformation, with values decreasing
from left to right)
along the horizontal axis.
The strength $G$ of the pairing interaction is given on
the far right, increasing
front to rear from $0.0$ to $0.3$ MeV.
For simplicity the proton and neutron pairing strengths
were set equal,
$G=G_\pi=G_\nu$.
}
\end{figure}
\noindent
\begin{figure}[b]
\caption{\label{fig:l4}
Spreading of the $L=4$ member of the ground band
over the 101 dimensional
$L=4$ \psma basis of the $[\,(fp)^{N_\pi=2}
(gds)^{N_\nu=2}\,]$ system,
see Fig. 3.
}
\end{figure}
\noindent
\begin{figure}[h]
\caption{\label{fig:l8}
Spreading of the $L=8$ member of the ground band
over the 101 dimensional
$L=8$ \psma basis of the $[\,(fp)^{N_\pi=2}
(gds)^{N_\nu=2}\,]$ system,
see Fig. 3.
}
\end{figure}
\noindent
\begin{figure}[h]
\caption{\label{fig:gsb}
Influence of the pairing strength (horizontal axis,
$G=G_\pi=G_\nu$) on members
of the ground state of the $[\,(fp)^{N_\pi=2}
(gds)^{N_\nu=2}\,]$ system.
Note that the rotational band on the left ($G=0$)
converts into a pairing
dominated level structure (seniority level scheme) for
($G=0.3 MeV$) on the
right.
A realistic range for the pairing strength is
approximately
($0.1 \mbox{MeV} \leq G \leq 0.2$ MeV).
}
\end{figure}
\noindent
\begin{figure}[htb]
\caption{\label{fig:mi}
Moments of inertia as a function of the pairing strength
($G=G_\pi=G_\nu$)
for yrast states of the $[\,(fp)^{N_\pi=2}
(gds)^{N_\nu=2}\,]$ system.
The moments of inertia $\theta$ were extracted from
the calculated spectrum,
see Fig. 6, by using the schematic formula $\theta =
\frac{L(L+1)}{2 E}$.
Values of the  angular momenta ($J=L$ since $S=0$)
are indicated on the right.
}
\end{figure}
\end{document}